\documentstyle[12pt,epsf]{article}

\begin{document}
\thispagestyle{empty}
\begin{flushright}
UTHEP-360\\
UT-CCP-P21\\
April 1997\\
\end{flushright}
\bigskip\bigskip\bigskip\begin{center}
{\LARGE {Domain wall fermions with}}
\vskip 7pt
{\LARGE {Majorana couplings}}
\end{center}  
\vskip 1.0truecm
\centerline{S. Aoki,} 
\vskip3mm
\centerline{Institute of Physics, University of Tsukuba}
\centerline{Tsukuba, Ibaraki 305, Japan}
\vskip5mm
\centerline{K. Nagai, and S. V. Zenkin\footnote{Permanent address: 
Institute for Nuclear 
Research of the Russian Academy of Sciences, 117312 Moscow, Russia.
E-mail address: zenkin@al20.inr.troitsk.ru}}
\vskip3mm
\centerline{Center for Computational Physics, University of Tsukuba} 
\centerline{Tsukuba, Ibaraki 305, Japan}
\vskip 1.0truecm
\bigskip \nopagebreak \begin{abstract}
\noindent

We examine the lattice boundary formulation of chiral fermions with either an
explicit Majorana mass or a Higgs-Majorana coupling introduced on one of the
boundaries. We demonstrate that the low-lying spectrum of the models with an
explicit Majorana mass of the order of an inverse lattice spacing is chiral at
tree level. Within a mean-field approximation we show that the systems with a
strong Higgs-Majorana coupling have a symmetric phase, in which a Majorana mass
of the order of an inverse lattice spacing is generated without spontaneous
breaking of the gauge symmetry. We argue, however, that the models within such
a phase have a chiral spectrum only in terms of the fermions that are singlets
under the gauge group. The application of such systems to nonperturbative
formulations of supersymmetric and chiral gauge theories is briefly discussed.

\end{abstract}

\newpage\setcounter{page}1
   
\section{Introduction}
On a lattice left-handed Weyl fermions $\psi$ are always accompanied by their
right-handed counterparts $\chi$, unless certain mild conditions for the action
are broken \cite{NN}. Such a doubling is present both in the Wilson \cite{KW}
and in the domain wall formulations \cite{DK,YS1} (for
a review, see \cite{KJ}), being the main obstacle
to the non-perturbative definition of chiral gauge theories. So far all
attempts to make the chiral counterpart $\chi$ sterile, in a way that does not
use a hard breaking of gauge symmetry and leaves an
interacting chiral theory, have failed\footnote{An exception is 
the class of formulations employing infinitely many fermionic degrees of 
freedom, either  
explicitly \cite{FS} or inexplicitly \cite{ov}; in the latter case  
the theory can not be formulated in terms of
the fermion action.} (see, for example, reviews
\cite{DP,KJ,YS2} and references therein).

Another conceivable way to define a theory with only fermion fields $\psi$ is
to decouple $\chi$ by giving them Majorana masses of the order of the
inverse lattice spacing. It can be done directly if the fermions belongs to
real representation of the gauge group. If they belong to complex
representation, a generalized Higgs mechanism is to be employed in order the
Majorana mass not to break the gauge invariance. In this case model must have
strong coupling paramagnetic (PMS) phase, where fermions acquire masses of the
order of the inverse lattice spacing, while the gauge symmetry is not 
spontaneously
broken, i.e. no chiral condensate or no vacuum expectation value of the Higgs
fields arise (for a review and further references, see \cite{PMS}).

In the Wilson formulation, the $\psi$ and $\chi$ are coupled through the Wilson
term. Therefore the introduction of a Majorana mass for $\chi$ generates a
Majorana coupling for $\psi$. So in the gauge theory fine tuning becomes
necessary not only for Dirac mass but also for Majorana mass of
$\psi$. Furthermore, in the case of complex representations, there arises a
serious problem with the properties of the model within the PMS phase, where
its spectrum either becomes vectorlike, or consists only of neutral,
i.e. singlet under the gauge group, chiral fermions, whose gauge interactions
very likely vanish in the continuum limit \cite{GPS,BDS,GPR1,AHK}. In the
domain wall formulation the chiral fermions $\psi$ and $\chi$ appear as
collective states of coupled five-dimension fermions. These states are
localized at two surfaces formed by mass defects in the five-dimensional system
\cite{DK} or by free boundaries of the five-dimensional space \cite{YS1}. These
surfaces are separated in the fifth dimension and the overlap between these
states is suppressed exponentially with this distance. This gives rise to a
hope that the above problems can be avoided in such a formulation.

This, in fact, is underlying idea of the recent proposal \cite{CEA} for lattice
formulation of the Standard Model. In order to generate the Majorana mass for
the collective state $\chi$, it has been suggested to introduce on the
surface at which $\chi$ is localized certain gauge invariant four-fermion
interactions motivated by the approach \cite{EP}. A similar idea is followed by
the proposal \cite{JN} for lattice formulation of $N=1$ supersymmetric
theories. In this case the Majorana mass is introduced on the same surface
directly, for the fermions belong to real representation.

Thus, yet more questions, which should be answered first, arise in such an
approach: ($i$) Whether the generation of the Majorana mass on one of the
surfaces leads to the chiral spectrum of the model? This question is common to
both proposals, \cite{CEA} and \cite{JN}, and requires special investigation,
since the chiral states in the domain wall formulation do not present in the
action explicitly. ($ii$) Whether the PMS phase exists in the systems employing
the Higgs mechanism? Although some indirect arguments in favour of the presence
of the PMS phase has already been given in \cite{CEA}, it seems interesting to
look at the problem from a more general point of view. ($iii$) Does the model
has chiral spectrum in the PMS phase, and if so, what fermions, charged or
neutral, form it? This is a crucial question to all formulations of chiral
gauge theories that employ the PMS phase.

The aim of this paper is to answer these questions. We consider the variant of
the domain wall formulation with free lattice boundaries \cite{YS1} and the
Majorana mass or Higgs-Majorana coupling introduced on one of the
boundaries. These models are introduced in Section 2. In Section 3 we analyze
the fermion propagators in the model with the Majorana mass and gauge
fields switched off, and show that the low-lying spectrum of such a model is
chiral. In Section 4 using a
mean-field technique we demonstrate the existence of the PMS phase in the
systems with the Higgs-Majorana coupling. In Section 5 we consider such systems
within the PMS phase and argue that they may have chiral spectrum only for the
fermions that are singlets under the gauge group. Section 6 contains a summary
and a discussion of possible applications of such models.

Our conventions are the following. We consider Euclidean hypercubic
$5$-dimensional lattice with spacing $a$, which is set to one unless otherwise
indicated, and volume $V = N^4 \times N_s$ with $N$ even. The lattice sites
numbered by $5$-dimensional vectors $(n, s) = (n_0, n_1, n_2, n_3, s)$, where
$n_{\mu} = -N/2+1, ..., N/2$, and $s = 0, ..., N_s$; $\hat{\mu}$ are unit
vectors along positive directions in four-dimensional space. Fermion (boson)
variables obey antiperiodic (periodic) boundary conditions for the first four
directions and free boundary conditions in the fifth direction.

\section{Model}

We consider the variant of the lattice domain wall fermions proposed in 
\cite{YS1}. The action of such a model can be written in the form:
\begin{eqnarray}
&&A_0[U] = \sum_{m, n, s, t}[\overline{\psi}_{m, s} \delta_{s t} 
D_{m n} \psi_{n, t} 
+ \overline{\chi}_{m, s} \delta_{s t}
\overline{D}_{m n} \chi_{n, t} \cr
&& \quad + \overline{\psi}_{m, s} (\delta_{s-1 \, t}\delta_{m n} - 
\delta_{s t} W_{m n})
\chi_{n, t} +
\overline{\chi}_{m, s} (\delta_{s+1 \, t}\delta_{m n} - 
\delta_{s t} W_{m n}) \psi_{n, t}],               \label{eq:DMF}
\end{eqnarray}
where $\psi_{n, s}$ and $\chi_{n, s}$ are two-component Weyl 
fermions in the five-dimensional space, transforming under the 
four-dimensional 
rotations as left- and right-handed spinors, respectively, 
\begin{eqnarray}
&&D = \nabla_0 + i \sum_{i} \sigma_{i} \nabla_{i}, \quad
\overline{D} = \nabla_0 - i \sum_{i} \sigma_{i} \nabla_{i}, \cr
&&\nabla_{\mu \; m n} = \frac{1}{2} (U_{m \: m+\hat{\mu}}
\delta_{m+\hat{\mu} \: n} - U_{m \: m-\hat{\mu}}
\delta_{m-\hat{\mu} \: n}), \cr
&&W_{m n} = \delta_{m n}( 1 - M) 
- \frac{1}{2}\Delta_{m n}, \cr
&&\Delta_{m n} = \sum_{\mu} (U_{m \: m+\hat{\mu}}
\delta_{m+\hat{\mu} \: n} + U_{m \: m-\hat{\mu}}
\delta_{m-\hat{\mu} \: n} - 2 \delta_{m n}),               \label{eq:def1}
\end{eqnarray}
$\sigma_i$ are the Pauli matrices, $U_{m \: m \pm \hat{\mu}}$ are  
four-dimensional gauge variables, and $M \in (0, 1)$\footnote{More precisely
the allowed mass range is $0 < M < 2$, 
but without loss of generality we can restrict it as indicated above.}
is intrinsic 
mass parameter of the formulation (`domain wall' mass). In (\ref{eq:DMF}) it
is implied that $\psi_{n, N_s+1} = \chi_{n, -1} = 0$. Both $\psi$ and $\chi$
belong to the same representation $g$ of the gauge group, so the action
(\ref{eq:DMF}) is gauge invariant. Without loss of 
generality we consider unitary groups.

The chiral fermions in such a formulation arise as surface modes: the left- 
and right-handed fermions are low momentum states localized at the $s = 0$ 
and $s = N_s$ boundaries, respectively. So these states are separated in 
the fifth dimension and overlap between them is suppressed 
exponentially with $N_s$.

The idea of refs. \cite{CEA,JN} is to introduce on the surface $s = N_s$
certain gauge invariant terms which might generate Majorana masses $O(1/a)$ for
right-handed fermions $\chi$. Such terms can always be represented in the form
bilinear in the fermions, so we consider the following action
\begin{eqnarray}
&&A[U, H] = A_0[U] + A_m[H], \cr
&&A_m[H] = \sum_{n}( \chi_{n, N_s}^{T}  
H_{n} \chi_{n, N_s} +
\overline{\chi}_{n, N_s} H^{\dagger}_{n} 
\overline{\chi}_{n, N_s}^{T} ).                   \label{eq:DMF1}
\end{eqnarray}

If representation $g$ is (pseudo)real, one can simply put
\begin{equation}
H_n = m \sigma_2,                                  \label{eq:mr}
\end{equation}
where $m$ is a certain constant symmetric matrix whose form  
depends on the group and ensures the gauge invariance of the mass 
term\footnote{Note that in the case of the pseudoreal 
groups constructing symmetric matrix $m$ requires more than one fermion 
generations.}. 

If $g$ belongs to complex representation, $H$ has the form
\begin{equation}
H_n = y \Phi_n \sigma_2,                            \label{eq:mc}
\end{equation}
where $y$ is the Higgs-Majorana coupling, and $\Phi$ is the Higgs field 
transforming under the gauge group as $\Phi_n \rightarrow g^{*}_{n} 
\Phi_n g^{\dagger}_n$. So the action (\ref{eq:DMF1}) is gauge invariant. 
In this case the Majorana mass $O(1/a)$ is to be generated without
spontaneous breaking of the gauge symmetry. It is achieved in the PMS phase, 
provided the system has such a phase. 

Let us examine first the case of the explicit mass (\ref{eq:mr}).

\section{Spectrum and propagators}

Complete information about Euclidean system is contained in the correlators, or
propagators, of the fields involving in the action. However, before we shall
analyse them, it seems to be instructive to get a qualitative idea about its
spectrum that can be easily obtained in a certain limiting case from the
equations of motion.

For our system with the gauge interactions switched off ($U = 1$) these
equations can be written in the momentum space as $L(p) \Psi(p) = 0$, where
$L(p)$ is $4 N_s \times 4 N_s$ matrix, and $\Psi$ is the column constructed
from the fields $\psi$, $\chi$, $\overline{\psi}^T$, and
$\overline{\chi}^T$. The solutions to this equations are determined by zeros of
det$L(p)$. Since the formulation is reflection positive, one can safely
continue the equations to the Minkowski space, so that $-p^2 \rightarrow
p_{0}^2 - p_i p_i$, where $p_i$ is three-vector. Then these zeros will
determine the energy spectrum of the system after its quantization.

To make the analysis as simple as possible, consider four dimensional space
continuous and take the limit $M \rightarrow 1$. Then $W = 1 - M \rightarrow
0$, and the above equation can be reduced to two independent sets of equations:
$L_{\psi} \psi = 0$, and $L_{\chi} \chi = 0$, where $L_{\psi, \chi}$ are
diagonal $N_s \times N_s$ matrices
\begin{eqnarray}
L_{\psi} = \mbox{diag}(p_{0}^2 - p_i p_i, \, p_{0}^2 - p_i p_i -1, \cdots
p_{0}^2 - p_i p_i -1, \, p_{0}^2 - p_i p_i -1), \qquad \quad \cr
L_{\chi} = \mbox{diag}(p_{0}^2 - p_i p_i - 1, \, p_{0}^2 - p_i p_i -1, \cdots
p_{0}^2 - p_i p_i -1, \, p_{0}^2 - p_i p_i - 4m^2).
\end{eqnarray}
From these expressions we immediately can read off that in the $\psi$ sector
there are one massless left-handed fermion ($E^2 = p_i p_i$) localized on the
surface $s = 0$, and $N_s$ massive solutions with the unit masses ($E^2 =
p_ip_i + 1$), that generally are not localized. In the $\chi$ sector there are
no massless fermions.  Instead, in addition to non-localized solutions with the
unit masses, which form with their $\psi$ counterparts $N_s$ massive Dirac
fermions, we have right-handed fermion with Majorana mass $2m$ localized on the
surface $s = N_s$. In the limit $m \rightarrow 0$ it turns to the mirror
massless mode of the domain wall formulation. Thus, we can conclude that such
simplified system has desirable chiral spectrum.

Of course, one can easily find also the fermion propagators in this 
limiting case. Returning to the Euclidean space we get
\begin{eqnarray}
&&\langle \psi_s \overline{\psi}_t\rangle = - \overline{D}
\left[\delta_{s t}\frac{1}
{p^2 + 1} 
+ \delta_{s 0} \delta_{t 0}\left(\frac{1}{p^2} 
- \frac{1}{p^2 + 1}\right)\right], \cr
&&\langle \chi_s \overline{\chi}_t\rangle = - D\left[\delta_{s t}
\frac{1}{p^2 + 1}
+ \delta_{s N_s} \delta_{t N_s}\left(\frac{1}{p^2 + 4 m^2} 
- \frac{1}{p^2 + 1}\right)\right], \cr
&&\langle \psi_s \overline{\chi}_t\rangle = \delta_{s \, t+1}\frac{1}
{p^2 + 1},
\quad s \neq 0, \, t \neq N_s, \cr
&&\langle \psi_0 \overline{\chi}_t\rangle =
\langle \psi_s \overline{\chi}_{N_s}\rangle = 0.
\end{eqnarray}

Let us now consider the fermion propagators of our lattice model. The most
simple way to do that is to express them in terms of the propagators determined
by the original action (\ref{eq:DMF}). We shall use for them notation $\langle
\Psi_A \overline \Psi_B \rangle_0$ with $\Psi_1 = \psi$ and $\Psi_2 =
\chi$. They have been calculated in \cite{YS1,AH} and in the four-dimensional
momentum space have the following form:
\begin{eqnarray}
&&\langle \psi\overline\psi \rangle_0 = -\overline D G_L \equiv - \overline D 
\frac{1}{\overline{p}^2 + W^{-}W^{+}}, \quad 
\langle \psi\overline\chi \rangle_0 = G_L W^{-} = W^{-}
G_R, \cr
&& \langle \chi\overline\chi \rangle_0 = - D G_R \equiv - D 
\frac{1}{\overline{p}^2 + W^{+}W^{-} }, \quad 
\langle \chi\overline\psi \rangle_0 = W^{+} G_L = G_R W^{+}, 
\label{eq:prop0}
\end{eqnarray}
where $W^{\pm}_{s \, t} = \delta_{s \pm 1 \, t} - \delta_{s t} W$, $W = 1 - M -
\Delta(p)/2$, and $\overline{p}_{\mu}$ and $\Delta(p)$ are Fourier transforms
of the operators $- i \nabla_{\mu}$ and $\Delta$ in (\ref{eq:def1}) at $U = 1$:
$\overline{p}_{\mu} = \sin p_{\mu}$, $\Delta(p) = 2 \sum_{\mu} (\cos p_{\mu} -
1)$. In the approximation specified bellow one has
\begin{eqnarray}
&&G_L(s,t) =
A_L e^{-\alpha (s+t)} + A_R e^{\alpha (s+t-2N_s)}
+ B e^{-\alpha |s-t|}, \cr
&&G_R(s,t) =
A_R e^{-\alpha (s+t)} + A_L e^{\alpha (s+t-2N_s)} + B e^{-\alpha |s-t|},
\label{eq:G}
\end{eqnarray}
with $\alpha = |\mbox{arccosh} \displaystyle \frac{1}{2}(W +
\frac{1+{\overline p}^2} W)|$. 
Here $A_L$, $A_R$ and $B$ are functions of $p$, but only $A_L$ has a pole
at $p^2 = 0$:
\begin{eqnarray}
A_L \rightarrow \frac{Z}{ \overline{p}^2 } - 2 + M,\quad A_R 
\rightarrow -\frac{(1-M)^2}{Z},
\quad B \rightarrow \frac{1}{Z}, \quad p^2 \rightarrow 0,
\label{eq:AB}
\end{eqnarray}
where $Z = M(2-M)$. Eqs. (\ref{eq:G}) and (\ref{eq:AB}) are given in the
approximation where all terms $O(e^{-\alpha N_s})$ are neglected. Since the
$\alpha$ is positive definite, the larger $N_s$, the better such an
approximation is justified.  At $p^2 \rightarrow 0$ one has $e^{-\alpha N_s}
\rightarrow (1-M)^{N_s}$, so this quantity can be considered as the accuracy
with which this formulation defines chiral fermions at finite $N_s$.

Now the propagators for the system (\ref{eq:DMF1}) with the Majorana mass
can be represented as
\begin{equation}
\langle \Psi_A \overline\Psi_B \rangle_m = \langle \Psi_A 
\exp \left(-A_m[\sigma_2 m]\right) \overline\Psi_B \rangle_0,
\end{equation}
and similarly for $\langle \Psi_A \Psi_B \rangle_m$ and 
$\langle \overline\Psi_A \overline\Psi_B\rangle_m$. Thus after some algebra we
arrive at the following expressions
\begin{eqnarray}
&&\langle\Psi_A \overline\Psi_B \rangle_m 
= \langle \Psi_A \overline \Psi_B \rangle_0 
-
4m^2 \langle \Psi_A \overline\chi (N_s)\rangle_0
\frac{\overline{D} G_R(N_s, N_s)}
{1 + 4m^2 \overline{p}^2 G_{R}^{2}(N_s, N_s)}
\langle \chi (N_s) \overline \Psi_B\rangle_0, \cr
&& \langle\Psi_A \Psi_B \rangle_m 
=-\langle \Psi_A \overline \chi (N_s)\rangle_0
\frac{2m \sigma_2}{1 + 4m^2 \overline{p}^2 G_{R}^{2}(N_s, N_s)}
\langle \overline\chi (N_s) \Psi_B \rangle_0, \cr
&& \langle\overline\Psi_A \overline\Psi_B \rangle_m 
=-\langle \overline\Psi_A \chi (N_s)\rangle_0
\frac{2m \sigma_2}{1 + 4m^2 \overline{p}^2 G_{R}^{2}(N_s, N_s)}
\langle \chi (N_s) \overline\Psi_B \rangle_0.              \label{eq:prop}
\end{eqnarray}

Consider now the low-momentum structure of the propagator $\langle\chi_s
\overline \chi_t \rangle_m$. In the approximation specified above we have: $G_R
(N_s,N_s) = (A_L+B)$, $G_R(s,N_s) = e^{\alpha (s-N_s)}(A_L+B)$, and $G_R(N_s,t)
= e^{\alpha (t-N_s)}(A_L+B)$. Then, from (\ref{eq:prop}) it follows that the
introduction of the Majorana mass modifies only function $A_L$
[cf. (\ref{eq:prop0}), (\ref{eq:G})]:
\begin{eqnarray}
&&\langle\chi_s \overline\chi_t \rangle_m 
= -D [ A_R e^{-\alpha (s+t)} 
+ A^m_L e^{\alpha (s+t-2N_s)} +  B e^{-\alpha |s-t|}], \cr
&&A^m_L = A_L - \frac{4m^2 \overline p^2 (A_L+B)^3}{1+4m^2 \overline p^2
(A_L+B)^2},
\end{eqnarray}
and this modification exactly cancels the pole in $A_{L}^{m}$:
\begin{eqnarray}
A_{L}^{m} \rightarrow \frac{1}{ 4m^2 Z } - \frac{1}{Z}, 
\quad p^2 \rightarrow 0.
\end{eqnarray} 
Note that this result matches well with our analysis in the beginning of this
section, for in the limit considered there $Z \rightarrow 1$.

On the other hand, in the same approximation the propagator 
$\langle\psi_s \overline \psi_t \rangle_m$ takes the form
\begin{eqnarray}
\langle\psi_s \overline\psi_t \rangle_m
= -\overline D [ B e^{-\alpha |s-t|} + A_L e^{-\alpha (s+t)} 
+ A^m_R e^{\alpha (s+t-2N_s)} ],
\end{eqnarray}
with
\begin{eqnarray}
A^m_R = A_R +  \frac{4m^2 (e^{-\alpha}-W)^2 (A_L+B)^3}
{1+4m^2 \overline p^2(A_L+B)^2}.
\end{eqnarray}
Since $\lim_{p^2\rightarrow 0} (e^{-\alpha}-W) = \displaystyle
- p^2 \frac{1 - M}{Z}$,
we get
\begin{eqnarray}
A^m_R \rightarrow A_R + \frac{(1-M)^2}{Z} = 0, \quad p^2 \rightarrow 0.
\end{eqnarray}
therefore, the pole structure of the propagator of the field $\psi$ is not 
affected by the Majorana mass acquired by the field $\chi$.
 
Finally, in order to see the effect of the fermion number violation in the
physical left-handed sector $\psi$, caused by the Majorana mass term for
$\chi$, we consider the propagator $\langle\psi_s \psi_t\rangle_m$. In the
above approximation we get
\begin{eqnarray}
&&\langle\psi_s \psi_t \rangle_m
= \frac{2m\sigma_2 (e^{-\alpha}-W)^2 (A_L+B)^2 }
{1+4m^2 \overline p^2(A_L+B)^2} \, e^{\alpha(s+t-2N_s)}\cr
&&\quad \rightarrow  \frac{\sigma_2 p^2}{2m Z^2} 
(1-M)^{(2N_s+2-s-t)},
\quad p^2 \rightarrow 0.
\end{eqnarray}
Thus, for physical left-handed fermions the effect is suppressed 
exponentially with $N_s$.

We should note that the action of the original domain wall model \cite{DK} can
be taken as the action $A_0$ in (\ref{eq:DMF}).  The analysis of the spectrum
and the propagators in this case, though is more complicated, leads essentially
to the same conclusions.

\section{Phase diagram}

Let us now consider the question of the existence of the PMS phase in the 
systems with the Higgs-Majorana coupling.

Whether a system has the PMS phase can be examined within mean-field 
approximation. We consider the case of radially frozen Higgs field, 
$\Phi^{\dagger}_{n} \Phi_{n} = 1$, using the technique developed in \cite{TZ}.
For that, it is sufficient to consider the gauge interactions turned off.

We write down the field $\Phi_n$ as $\Phi_n = \sum_a T^a \phi^{a}_{n}$, with
$\sum_a (\phi^{a}_{n})^{2} = 1$, and $T^a$ matrices determined by the
group and its representation. Then the action for the $\Phi$ takes the form 
\begin{equation}
A_{\Phi} = - 2 \kappa \sum_{m, \mu, a} \phi^{a}_{n} \phi^{a}_{n + \hat{\mu}},
\end{equation}
where $\kappa$ is a hopping parameter. Not to break the reflection positivity
of the model, we shall consider it at $\kappa > 0$. The partition function of 
the system reads as
\begin{eqnarray}
&&Z = \int \prod_n d \Phi_n \prod_s d\psi_{n,s} d\overline{\psi}_{n,s}  
d\chi_{n,s} d\overline{\chi}_{n,s} 
\exp (-A_{\Phi} -  A[1, y \Phi \sigma_2]) \cr
&&\quad = \int \prod_n d \Phi_n \exp (-A_{\Phi} + \ln Z_f[y \Phi]),
\label{eq:sys1}
\end{eqnarray}
where $d \Phi$ is the Haar measure on the group, and $Z_f[y \Phi]$ is the 
fermion partition function in the external field $\Phi$.

According to \cite{TZ}, the critical lines separating symmetric
(paramagnetic) phases, where the vacuum expectation value of the Higgs 
field $\langle \Phi \rangle = 0$, from the broken (ferromagnetic) 
phases, where  $\langle \Phi \rangle \neq 0$, are determined by
the expression
\begin{equation}
\left. \kappa_{cr}(y) = \frac{c_1}{8} - \frac{c_2}{4} \frac{\partial^2}
{\partial h^2}
\lim_{N \rightarrow \infty} \frac{1}{N^4} \langle \ln Z_f[y \Phi] 
\rangle_h \right|
_{h = 0},                                               \label{eq:kkr}
\end{equation}
where $h = [\sum_a (h^a)^2]^{1/2}$ is the mean field, 
\begin{equation}
\langle \ln Z_f[y \Phi] \rangle_h = \frac{\int \prod_n d \Phi_n 
\ln Z_f[y \Phi]
\exp \sum_{n, a} h^a \phi^{a}_{n}}{\int \prod_n d \Phi_n 
\exp \sum_{n, a} h^a \phi^{a}_{n}},                         \label{eq:exv}
\end{equation}
and $c_1$ and $c_2$ are group dependent positive numbers. For 
instance, for U(1) one has $c_1 = c_2 = 1$.

Integrating in (\ref{eq:sys1}) over $\psi$, and then over $\chi$ 
successively slice by slice in the fifth direction, we find
\begin{equation}
Z_f[y \Phi] = \mbox{Det}^{N_s+1}[D] \: \prod_{s = 0}^{N_s-1} 
\mbox{Det}[B_s] \: 
\mbox{Det}^{1/2} [4 y^2 + \Phi^{\dagger} \sigma_2 B_{N_s}^{T} 
\Phi \sigma_2 B_{N_s}],                            \label{eq:z1}
\end{equation}
where Det means the determinant in the space of four-dimensional lattice and 
spinors indices, and the operators $B_s$ are determined by the 
recursive relations
\begin{eqnarray}
&&B_{N_s} = \widetilde{D} + D^{-1} - \frac{C^2}{B_{N_s-1}}, \cr
&&B_s = \widetilde{D} - \frac{C^2}{B_{s-1}}, \quad 0 < s <
N_{s}, \cr
&&B_0 = \widetilde{D},                                  \label{eq:rea}
\end{eqnarray} 
with $\widetilde{D} = \overline{D} - D^{-1} -W^2 D^{-1}$, and
$C = - W D^{-1}$.

Since one is interested in the paramagnetic phase at strong coupling $y$,
it is sufficient to know expectation value (\ref{eq:exv}) in the leading 
order in 
$y^{-2}$ and $h^2$. From (\ref{eq:exv}), (\ref{eq:z1}), and (\ref{eq:rea}), 
using formulae of ref. \cite{TZ}, we get
\begin{equation}
\langle \ln Z_f[y \Phi] \rangle_h = - \frac{c_3}{32 y^2} h^2 
\mbox{Tr} [\sigma_2 B_{N_s}^{T} \sigma_2 B_{N_s}]
+ O(h^0) + O(\frac{1}{y^4}, h^3),                   \label{eq:exp2}
\end{equation}
where Tr means the trace over the spatial and spinor indices, and $c_3$  
is group dependent positive number; in the case of U(1) it is equal to 
one, as well.

Finally, from (\ref{eq:kkr}), (\ref{eq:rea}), and (\ref{eq:exp2}), we
get:
\begin{eqnarray}
&&\kappa_{cr}(y) = \frac{c_1}{8} - \frac{c_2 c_3}{32 y^2} I(N_s, M), \cr
&&I(N_s, M) = \int_{{\cal B}} \frac{d^4 p}{(2 \pi)^4} 
\frac{1}{\overline{p}^2}
E^2(p, N_s),
\end{eqnarray}
where 
\begin{equation}
E(p, N_s) = 1 + F_0(p) - \frac{W^2(p)}
{F_1(p) - \frac{\textstyle W^2(p)}
 {\textstyle F_2(p) - \frac{\textstyle W^2(p)}
{\frac{\vdots }
  {\textstyle F_{N_s-1}(p) - \frac{\textstyle W^2(p)}
{\textstyle F_{N_s}(p)}}}}},
\end{equation}
and $F_s(p) = - 1 - \overline{p}^2 - W^2(p)$. 
We find $I(N_s, M)$ numerically practically independent of $N_s$. For
instance at $M = 0.8$ we have: $I(1, 0.8) = 376.4\cdots$, $I(5, 0.8) = 
375.6\cdots$, $I(20, 0.8) = 375.6\cdots$. 

Thus, the system is in the PMS phase at $y > y_I = (c_2 c_3 I/ c_1)^{1/2}/2
= O(10)$, 
and $\kappa < \kappa_{cr}(y)$.  
The corresponding phase diagram is shown in 
Fig. 1. 
\begin{figure}[htb]
   \vspace*{0cm}
   \epsfysize=7cm
   \epsfxsize=10cm
   \centerline{\epsffile{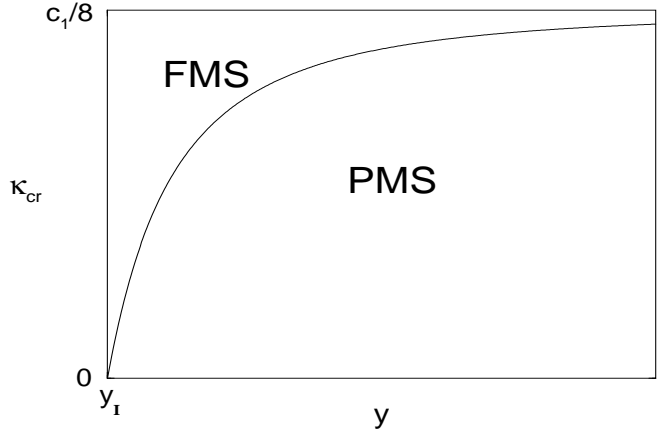}}
   \vspace*{-0cm}
\caption{The generic phase diagram for the system (3) at strong Higgs-Majorana
coupling; FMS and PMS are the strong coupling ferromagnetic (broken) and
paramagnetic (symmetric) phases, respectively.}
\end{figure}
This confirms the indirect arguments of ref. \cite{CEA}. 
 
Let us assume now that the system is within such a phase and consider what may
happen to it there.

\section{Within the PMS phase}

Since the Higgs field fluctuates strongly in this phase, and $\langle \Phi
\rangle = 0$, standard methods of examination of such systems, working in 
the broken phases, is inapplicable here. We however still can get some idea of
what may happen in this phase making use the method proposed in \cite{JS,ALS}.

Let us represent the Higgs field in (\ref{eq:DMF1}) as 
\begin{equation}
\Phi_n = \widetilde{\Phi}_{n}^{T} \widetilde{\Phi}_n, 
\end{equation}
where the group valued field $\widetilde
{\Phi}$, $\widetilde{\Phi}_{n}^{\dagger} \widetilde {\Phi}_n = 1$, transforms
under the gauge group as $\widetilde{\Phi}_n \rightarrow
\widetilde{\Phi}_n g_{n}^{\dagger}$. Following
the arguments of \cite{JS,ALS}, we assume that adequate nonzero parameter in
the PMS phase is the link expectation value
\begin{eqnarray}
z^2 = \langle \widetilde{\Phi}^{\dagger}_{n} \widetilde{\Phi}_{n \pm 
\hat{\mu}}\rangle, \quad 
\langle\widetilde{\Phi}_{n}\rangle = \langle\widetilde{\Phi}^{\dagger}_{n}
\rangle = 0.                                              \label{eq:vev}
\end{eqnarray}

Further results will depend on the dynamical variables in terms of which the
consideration is performed. Therefore we consider all the three possibilities.

{\bf 1.} Introduce the gauge group singlet fields 
\begin{equation}
\widetilde{\chi}_{N_s} = z \widetilde{\Phi} \chi_{N_s}, \quad 
\overline{\widetilde{\chi}}_{N_s} = 
z \overline{\chi}_{N_s}\widetilde{\Phi}^{\dagger}.
                                                       \label{eq:chin}
\end{equation}
Then, with taking into account (\ref{eq:vev}), the system (\ref{eq:DMF1}) 
in the PMS phase at tree level can be described by the action
\begin{eqnarray}
&&\widetilde{A}_1= \sum_{s, t \neq N_s}(\overline{\psi}_{s} 
\delta_{s t} D \psi_{t} + \overline{\chi}_{s} \delta_{s t}
\overline{D}\chi_{t}  
+ \overline{\psi}_{s} W^{-}_{s t}\chi_{t} 
+ \overline{\chi}_{s} W^{+}_{s t} \psi_{t}) \cr
&& \qquad \qquad+ \overline{\psi}_{N_s} D {\psi}_{N_s} 
+ \overline{\psi}_{N_s} {\chi}_{N_s-1}
+ \overline{\chi}_{N_s-1} {\psi}_{N_s} \cr 
&& \qquad \qquad            
+ \overline{\widetilde{\chi}}_{N_s} \overline{D} \widetilde{\chi}_{N_s} 
+ \frac{y}{z^2} (\widetilde{\chi}_{N_s}^{T} \sigma_2 \widetilde{\chi}_{N_s} 
+ \overline{\widetilde{\chi}}_{N_s} \sigma_2 
\overline{\widetilde{\chi}}_{N_s}^{T}),
                                                        \label{eq:ns1}
\end{eqnarray}
where we omit summation over four-dimensional indices. We see that in this
case the massive neutral fermion $\widetilde{\chi}_{N_s}$ is decoupled, for the
Wilson term at $s = N_s$ vanishes. This made $\widetilde{\chi}_{N_s}$ and
$\psi_{N_s}$ naive, and although at the first glance this should not affect the
chiral properties of the system, since both these fermions still have masses of
the order of the cutoff, it turns out that it gives rise to species doubling
of a massless mode. It is clearly seen from the structure of the fermion
determinant, which in this case takes the form [cf. (\ref{eq:z1})]
\begin{equation}
Z_1 = \mbox{Det}^{N_s+1}[D] \: \prod_{s = 0}^{N_s-1} 
\mbox{Det}[B_s] \: 
\mbox{Det}^{1/2} [4 \frac{y^2}{z^4} - D \overline{D}],         \label{eq:zn1}
\end{equation}
where the functions $B_s$ are determined in (\ref{eq:rea}). Since $B_s \propto
D^{-1}$, the determinant has zero at the corners of the
Brillouine zone. So this scenario leads to the failure of the model. 

{\bf 2.} Together with the neutral fields (\ref{eq:chin})
one can also introduce the neutral fields
\begin{equation}
\widetilde{\psi}_{N_s} = z \widetilde{\Phi} \psi_{N_s}, \quad 
\overline{\widetilde{\psi}}_{N_s} = 
z \overline{\psi}_{N_s}\widetilde{\Phi}^{\dagger}.
                                                       \label{eq:psin}
\end{equation}
In terms of these variables the system in the PMS phase takes the form
\begin{eqnarray}
&&\widetilde{A}_2= \sum_{s, t \neq N_s}(\overline{\psi}_{s} 
\delta_{s t} D \psi_{t} + \overline{\chi}_{s} \delta_{s t}
\overline{D}\chi_{t}  
+ \overline{\psi}_{s} W^{-}_{s t}\chi_{t} 
+ \overline{\chi}_{s} W^{+}_{s t} \psi_{t}) \cr
&& \qquad \qquad+ \overline{\widetilde\psi}_{N_s} D {\widetilde\psi}_{N_s} 
+ \overline{\widetilde{\chi}}_{N_s} \overline{D} \widetilde{\chi}_{N_s} 
- \overline{\widetilde\psi}_{N_s}\widetilde{W}\widetilde{\chi}_{N_s}           
- \overline{\widetilde\chi}_{N_s}\widetilde{W}\widetilde{\psi}_{N_s} \cr
&& \qquad \qquad 
+ \frac{y}{z^2} (\widetilde{\chi}_{N_s}^{T} \sigma_2 \widetilde{\chi}_{N_s} 
+ \overline{\widetilde{\chi}}_{N_s} \sigma_2 
\overline{\widetilde{\chi}}_{N_s}^{T}),
                                                        \label{eq:ns2}
\end{eqnarray}
where in the four-dimensional momentum space 
\begin{eqnarray}
\widetilde{W} = \frac{1}{z^2}(1-M) - \sum_{\mu}\left(\cos p_{\mu} 
- \frac{1}{z^2}\right).                                  \label{eq:W2}
\end{eqnarray}
In this case the slice $s = N_s$ decouples from the rest of the lattice, so
that one has massive neutral fermions at $s = N_s$ and the original massless
model on the rest of the lattice. The neutral fermions
$\widetilde{\psi}_{N_s}$, $\widetilde{\chi}_{N_s}$ can still be made massless,
by setting the domain wall mass to $M = 5 - 4 z^2$, but this does not rescue
the situation with the rest of the lattice. So in this case the model fails,
too.

{\bf 3.} Finally, in terms of the singlet fields like (\ref{eq:chin}) and
(\ref{eq:psin}) introduced at all $s$, the action reads as
\begin{eqnarray}
&&\widetilde{A}_3 = \sum_{s t}(\overline{\widetilde{\psi}}_s \delta_{s t}
D \widetilde{\psi}_t 
+ \overline{\widetilde{\chi}}_s\delta_{s t} \overline{D} \widetilde{\chi}_t 
+ \overline{\widetilde{\psi}}_s \widetilde{W}^{-}_{s t} \widetilde{\chi}_{t} 
+ \overline{\widetilde{\chi}}_s \widetilde{W}^{+}_{s t} \widetilde{\psi}_t) \cr
&&\qquad + \frac{y}{z^2} (\widetilde{\chi}_{N_s}^{T} \sigma_2 
\widetilde{\chi}_{N_s} 
+ \overline{\widetilde{\chi}}_{N_s} \sigma_2 
\overline{\widetilde{\chi}}_{N_s}^{T}),
                                                        \label{eq:ns3}
\end{eqnarray}
where 
\begin{eqnarray}
\widetilde{W}^{\pm}_{s t} = \frac{1}{z^2} \delta_{s \pm 1 \, t} 
-\delta_{s t}\widetilde{W}.
                                                              \label{eq:W3}
\end{eqnarray}
 
We see that these system is equivalent to that considered in the previous
section, only if $z = 1$. However one can hardly expect that it is so on the
ground of the estimations of $z^2$ in pure scalar models \cite{BDS,AHK}. For
instance, in $O(2)$ and $O(4)$ models it has been found $z^2 \simeq 0.19$.
Therefore our previous analysis is not directly applicable to this system.
However we can made some qualitative conclusions about the effects of such
renormalization.

An analysis similar to that of the beginning of the previous section, shows
that in the continuum case, when the Wilson term is equal to zero, $z$ simply
renormalizes the masses of the massive excitations: now they become $1/z^2$,
rather than 1. The situation on the lattice is more complicated. Indeed, making
the small $p$ expansion in (\ref{eq:W3}) one can see that the role of the
factor $1-M$, which provides the localization of the massless states at the
boundaries, now is played by $z$ dependent combination $5-4z^2 -M$, that may
considerably shift the range of admissible values of $M$.  This requires a
priori knowledge of the $z$, that complicates considerably the numerical study
of such models.  But the main problem in this variant is that all the fermions
in the system now are neutral.

The question of which of these scenarios is realized actually in the system
requires special investigation, and the answer may depend on the strength of
the Wilson parameter $r$, set to 1 in this paper, as well as on the couplings
$y$ and $\kappa$.  However none of them leads to the chiral spectrum of
charged, i.e. gauge non-singlet, fermions which one tends to describe. So we
can conclude that in this respect the domain wall fermions have no appreciable
advantages compared with the four dimensional Wilson fermions.

\section{Summary and prospects}

The lattice boundary formulation of chiral fermions with Majorana mass
introduced for the field $\chi$ on the boundary $s = N_s$ indeed yields the
low-lying chiral spectrum at the tree level: there exists only one massless
left-handed fermion $\psi$ localized at the boundary $s = 0$. Our results also
show that effects caused by the introduction of such Majorana mass is strongly
suppressed in physical sector. For instance, the fermion number violation for
the $\psi$ is suppressed exponentially as the size of the fifth dimension $N_s$
increases.

Since the Majorana mass does not break the gauge invariance for real
representations of the gauge group, the most immediate implementation of such
systems is the lattice formulation of the $N=1$ supersymetric models 
\cite{JN}. Our results speaks in favour of this idea, since the exponential
suppression of the undesirable effects in the physical sector is an indication
that the problem of fine tuning can be avoided in such a formulation, at least
in the perturbation theory. This however should be demonstrated explicitly,
and this work now is in progress.

The prospect for chiral gauge theories is less certain, for they have to employ
the generalized Higgs mechanism in the PMS phase. Although we have demonstrated
that the PMS phase can exist in these models, their possible dynamics within
such a phase does not give a ground for optimism.  Indeed, we argued that the
spectrum of charge fermions within the PMS phase is vectorlike, and that only
spectrum of the neutral states, by a certain tuning of the domain wall mass,
can be made chiral. Thus the crucial question is what is the gauge
interactions of such neutral chiral states. Previous studies of similar states
appearing in the models with the Wilson-Yukawa couplings give a strong evidence
that such states in the continuum limit become non-interacting
\cite{GPS,BDS,GPR1,AHK}. It is this point that leads to very plausible failure
of those models, as well as the models with multifermion couplings 
\cite{GPR2}. Such screening of the chiral charges appears to be an universal
phenomenon pursuing any models within the PMS phase. Therefore we consider this
point as the main problem on the way of implementation of the domain wall
fermions to exactly gauge invariant nonperturbative formulation of chiral gauge
theories.\\

\vskip10pt

\noindent{\Large \bf Acknowledgements}\\

S.V.Z. is grateful to the staff of the Center for Computational Physics, where
this work has been done, and particularly to Y. Iwasaki and A. Ukawa, for their
kind hospitality.  S.A. was supported in part by the Grants-in-Aid of the
Ministry of Education (No.08640350); S.V.Z. was supported by ``COE Foreign
Researcher Program'' of the Ministry of Education of Japan and by the Russian
Basic Research Fund under the grant 95-02-03868a.

\end{document}